\documentclass[12pt]{article}

\newcommand{\eps}{\varepsilon}

\newcommand{\om}{\omega}

\begin{document}

\title{Polariton effect in nonlinear pulse propagation}

\author{S.A. Darmanyan$^{(1)}$, A.M. Kamchatnov$^{(1)}$,
and M. Nevi\`{e}re$^{(2)}$\\
$^{(1)}$Institute of Spectroscopy, Russian Academy of Sciences,\\ Troitsk,
Moscow Region 142190, Russia\\
$^{(2)}$Institut Fresnel, Facult\'{e} des Sciences et Techniques
de Saint J\'{e}rome,\\ case 262, 13397 Marseille Cedex 20, France}

\maketitle


\begin{abstract}
The joint influence of the polariton effect and Kerr-like nonlinearity on
propagation of optical pulses is studied. The existence of different
families of envelope solitary wave solutions in the vicinity of polariton
gap is shown. The properties of solutions depend strongly on the
carrier wave frequency. In particular, solitary waves inside and outside of
the polariton gap exhibit different velocity and amplitude dependencies on
their duration.
\end{abstract}

pacs{42.65.Tg,  42.25 Bs}


\section{Introduction}

During the last years, due to fast progress in the fabrication of
microresonators, organic and inorganic quantum wells etc., great interest
has been attracted to investigation of electromagnetic properties of
these new objects, including propagation of nonlinear pulses in such
structures (see, e.g., \cite{BW}-\cite{AKBW}). For example, in \cite{AKBW}
propagation of nonlinear pulses along a quantum well imbedded in a
microresonator was studied for two types of nonlinearities---Kerr-like
nonlinearity applied to envelopes of long enough pulses, and self-induced
transparency (SIT) nonlinearity applied to short and intense pulses with
frequency close the two-level resonance. In this study, the authors have chosen
such conditions of propagation that one can neglect the polariton effect of
formation of a gap in the dispersion law of an electromagnetic wave coupled
with the wave of polarization in a medium. Then the problem of pulse
propagation can be reduced to either nonlinear Schr\"{o}dinger (NLS)
equation, or sine-Gordon equation with well-known soliton solutions.
However, the region of frequencies in vicinity of the polariton gap is
very important, because some properties of the structures under
consideration show up in this region only.

One should note that related problems were already studied long time ago in
the theory on nonlinear pulse propagation through a medium in vicinity
of exciton resonances. In papers \cite{SH}, the polariton SIT pulses were
found, but later there was made a statement \cite{Han,In} that polariton
effect prevents the existence of the SIT pulses. This contradiction was
clarified in Refs.~\cite{AI}, where it was shown that polariton solitons do
exist due to subtle balance of small effects, and therefore these solutions
may easily be overlooked if too rude approximation is made in the
evolution equations. Later the authors of \cite{BMB,GH} confirmed in general this
result and described some additional remarkable properties of the polariton
SIT pulses in the vicinity of polariton gap beyond the perturbation theory.
Similar problems have also been
studied for the case of Kerr nonlinearity (see, e.g., Ref.~\cite{Mosk}
and references therein). However, the approximations used were not justified
well enough, and some properties of polariton solitons remained
unclear. In paper \cite{TCA}, for long pulses and frequency of the carrier
wave far enough from the polariton gap, the governing equations were reduced
to the perturbed NLS equation for the envelope and corresponding soliton
solutions are described. A closely related problem, namely a pulse propagation
in Kerr-nonlinear medium with singular dispersion relation was studied in
\cite{PPL} where bright and dark solitary wave solutions were found in
vicinity of linear resonance.

In this paper we perform a thorough investigation of polariton solitons for
the case of Kerr nonlinearity following the method developed by Akimoto and
Ikeda \cite{AI} and show the existence of localized solutions both inside
and outside the polariton gap.

\section{Main equations}

We start with the standard equations of the classical theory of
electromagnetic waves propagating through an isotropic medium (see, e.g.,
\cite{AG}),
\begin{equation}
\frac{\partial ^{2}\mathbf{E}}{\partial x^{2}}-\frac{\epsilon _{0}}{c^{2}}%
\frac{\partial ^{2}\mathbf{E}}{\partial t^{2}}=\frac{4\pi }{c^{2}}\frac{%
\partial ^{2}\mathbf{P}}{\partial t^{2}},  \label{eq1}
\end{equation}
\begin{equation}
\frac{\partial ^{2}\mathbf{P}}{\partial t^{2}}+\omega _{T}^{2}\mathbf{P}%
+\chi \left| \mathbf{P}\right| ^{2}\mathbf{P}=\alpha \mathbf{E},  \label{eq2}
\end{equation}
where $\epsilon _{0}$ is the background dielectric constant, and
\begin{equation}
\alpha =\epsilon _{0}(\omega _{L}^{2}-\omega _{T}^{2})/4\pi .  \label{eq3}
\end{equation}
These equations describe interaction of the electromagnetic field $\mathbf{E}$
and the polarization wave $\mathbf{P}$ due to the
Kerr-like nonlinearity (measured by the parameter $\chi $). Here we ignore
the effect of damping. The parameters, $\omega _{T}^{2}$ and $\omega
_{L}^{2} $ characterize the dispersion law $\omega =\omega (k)$ of linear
waves, which is given by the equation
\begin{equation}
F(k,\omega )\equiv \omega _{T}^{2}-\omega ^{2}-\frac{4\pi }{c^{2}}\frac{%
\alpha \omega ^{2}}{k^{2}-\epsilon _{0}\omega ^{2}/c^{2}}=0.  \label{eq4}
\end{equation}
As follows from (\ref{eq4}) the dispersion law has a gap in the frequency
interval
\begin{equation}
\omega _{T}<\omega <\omega _{L},  \label{eq5}
\end{equation}
where linear waves cannot propagate. As was mentioned above for frequency
far enough from the polariton gap (\ref{eq5}), one can introduce the
envelope function and the system (\ref{eq1}), (\ref{eq2}) can be reduced to
the NLS equation possessing well-known soliton solutions. Here, we are
interested in solutions of the system (\ref{eq1}), (\ref{eq2}) for frequency
near and inside the polariton gap (\ref{eq5}).

We look for the solutions in the form of stationary linearly polarized
waves, so that $\mathbf{E}$ and $\mathbf{P}$ may be considered as the
following scalar functions
\begin{equation}
E(x,t)=\mathcal{E}(t-x/V)e^{i\theta },\quad P(x,t)=\left[ u(t-x/V)-iv(t-x/V)%
\right] e^{i\theta },  \label{eq6}
\end{equation}
where $V$ is velocity of the pulse, and the phase $\theta (x,t)$ is
\begin{equation}
\theta (x,t)=kx-\omega t-\phi (t-x/V).  \label{eq7}
\end{equation}
Substitution of Eqs.~(\ref{eq6}), (\ref{eq7}) into Eqs.~(\ref{eq1}), (\ref
{eq2}) leads to the following system of equations for the variables $u,\,v,\,%
\mathcal{E},\,\phi $:
\begin{equation}
\ddot{u}-\left( \omega ^{2}-\omega _{T}^{2}+2\omega \dot{\phi}+{\dot{\phi}}%
^{2}\right) u-2(\omega +\dot{\phi})\dot{v}-{\ddot{\phi}}v+\chi
(u^{2}+v^{2})u=\alpha \mathcal{E},  \label{eq8}
\end{equation}
\begin{equation}
\ddot{v}-\left( \omega ^{2}-\omega _{T}^{2}+2\omega \dot{\phi}+{\dot{\phi}}%
^{2}\right) v+2(\omega +\dot{\phi})\dot{u}+{\ddot{\phi}}u+\chi
(u^{2}+v^{2})v=0,  \label{eq9}
\end{equation}
\begin{equation}
\begin{array}{ll}
\left( \frac{1}{V^{2}}-\frac{\epsilon _{0}}{c^{2}}\right) \ddot{\mathcal{E}}%
& -\left[ \left( k^{2}-\frac{\epsilon _{0}}{c^{2}}\omega ^{2}\right)
+2\left( \frac{k}{V}-\frac{\epsilon _{0}\omega }{c^{2}}\right) \dot{\phi}%
+\left( \frac{1}{V^{2}}-\frac{\epsilon _{0}}{c^{2}}\right) {\dot{\phi}}^{2}%
\right] \mathcal{E} \\
& =(4\pi /c^{2})\left[ \ddot{u}-u(\omega +\dot{\phi})^{2}-\ddot{\phi}%
v-2(\omega +\dot{\phi})\dot{v}\right] ,
\end{array}
\label{eq10}
\end{equation}
\begin{equation}
\begin{array}{ll}
\left( \frac{1}{V^{2}}-\frac{\epsilon _{0}}{c^{2}}\right) \ddot{\phi}%
\mathcal{E}& +2\left[ \frac{k}{V}-\frac{\epsilon _{0}\omega }{c^{2}}+\left(
\frac{1}{V^{2}}-\frac{\epsilon _{0}}{c^{2}}\right) \dot{\phi}\right] \dot{%
\mathcal{E}} \\
& =(4\pi /c^{2})\left[ \ddot{v}-v(\omega +\dot{\phi})^{2}+\ddot{\phi}%
u+2(\omega +\dot{\phi})\dot{u}\right] ,
\end{array}
\label{eq11}
\end{equation}
where the overdot stands for the derivative with respect to $\xi =t-x/V$.

\section{Linear approximation}

We suppose that at the tails of the pulse (i.e. at infinite $|\xi |$) the
variables $u,\,v,\,\mathcal{E},\,\phi $ go to zero, so that in these regions
the system (\ref{eq8})--(\ref{eq11}) can be linearized:
\begin{equation}
\ddot{u}-\left( \omega ^{2}-\omega _{T}^{2}\right) u-2\omega \dot{v}=\alpha
\mathcal{E},  \label{eq12}
\end{equation}
\begin{equation}
\ddot{v}-\left( \omega ^{2}-\omega _{T}^{2}\right) v+2\omega \dot{u}=0,
\label{eq13}
\end{equation}
\begin{equation}
\left( \frac{1}{V^{2}}-\frac{\epsilon _{0}}{c^{2}}\right) \ddot{\mathcal{E}}%
-\left( k^{2}-\frac{\epsilon _{0}}{c^{2}}\omega ^{2}\right) \mathcal{E}%
=(4\pi /c^{2})\left( \ddot{u}-\omega ^{2}u-2\omega \dot{v}\right) ,
\label{eq14}
\end{equation}
\begin{equation}
2\left( \frac{k}{V}-\frac{\epsilon _{0}\omega }{c^{2}}\right) \dot{\mathcal{E%
}}=(4\pi /c^{2})\left( \ddot{v}-\omega ^{2}v+2\omega \dot{u}\right) .
\label{eq15}
\end{equation}
For the exponential dependence
\begin{equation}
\left( \mathcal{E},u,v\right) =\left( \mathcal{E}_{0},u_{0},v_{0}\right)
\exp (-|\xi |/\tau ),\qquad \mbox{\rm at\ \ }|\xi |\rightarrow \infty ,
\label{eq16}
\end{equation}
where $\tau $ is the duration of the pulse, the system (\ref{eq12})--(\ref
{eq15}) reduces to algebraic equations which define the ``dispersion law''
and the velocity of the pulse as functions of $\tau $. It is convenient to
introduce the variables
\begin{equation}
X=ck/\omega ,\qquad Y=c/V,  \label{eq17}
\end{equation}
and to define the characteristic parameters
\begin{equation}\label{eq18}
s=1/\omega \tau ,\qquad \Lambda ^{2}=1/(\omega _{L}^{2}-\omega _{T}^{2})\tau
^{2},
\end{equation}
so that the ratio
\begin{equation}
\frac{\Lambda ^{2}}{s^{2}}=\frac{\omega ^{2}}{\omega _{L}^{2}-\omega _{T}^{2}%
}\equiv \Omega ^{2}  \label{eq19}
\end{equation}
does not depend on $\tau $. The variable
\begin{equation}
\Delta ^{2}=\frac{\omega ^{2}-\omega _{T}^{2}}{\omega _{L}^{2}-\omega
_{T}^{2}}  \label{eq20}
\end{equation}
measures the frequency in vicinity of the polariton gap. The cases of $%
\Delta ^{2}>1$ and $\Delta ^{2}<0$ correspond to upper and lower polariton
branches respectively. By substituting Eq.~(\ref{eq16}) into Eqs.~(\ref{eq12}%
)--(\ref{eq15}), we arrive at the system
\begin{equation}
\begin{array}{ll}
& \left( \Lambda ^{2}-\Delta ^{2}\right) u_{0}-2s\Omega ^{2}v_{0}=(\epsilon
_{0}/4\pi )\mathcal{E}_{0}, \\
& \left( \Lambda ^{2}-\Delta ^{2}\right) v_{0}+2s\Omega ^{2}u_{0}=0, \\
& \left( X^{2}-s^{2}Y^{2}\right) \mathcal{E}_{0}=\epsilon _{0}(1-s^{2})%
\mathcal{E}_{0}+4\pi \lbrack (1-s^{2})u_{0}+2sv_{0}], \\
& 2s\left( XY-\epsilon _{0}\right) \mathcal{E}_{0}=4\pi \lbrack
(1-s^{2})v_{0}-2su_{0}].
\end{array}
\label{eq21}
\end{equation}

Eliminating $u_{0}$ and $v_{0}$ from Eqs.~(\ref{eq21}) gives the following\
system
\begin{equation}
\begin{array}{ll}
X^{2}-s^{2}Y^{2}& =\epsilon _{0}\left[ 1-s^{2}+\frac{(1-s^{2})(\Lambda
^{2}-\Delta ^{2})-4s^{2}\Omega ^{2}}{(\Lambda ^{2}-\Delta
^{2})^{2}+4s^{2}\Omega ^{4}}\right] , \\
XY& =\epsilon _{0}\left[ 1+\frac{\Lambda ^{2}-\Delta ^{2}+(1-s^{2})\Omega
^{2}}{(\Lambda ^{2}-\Delta ^{2})^{2}+4s^{2}\Omega ^{4}}\right] ,
\end{array}
\label{eq22}
\end{equation}
the solution of which with account of Eq.~(\ref{eq19}) yields
\begin{equation}
\begin{array}{ll}
X^{2}& =\left( \frac{ck}{\omega }\right) ^{2}=\frac{\epsilon _{0}}{2}\Big[%
(1-s^{2})-\frac{(1-s^{2})(\Delta ^{2}-\Lambda ^{2})+4\Lambda ^{2}}{(\Delta
^{2}-\Lambda ^{2})^{2}+4\Lambda ^{2}\Omega ^{2}} \\
& +(1+s^{2})\sqrt{\frac{(\Delta ^{2}-\Lambda ^{2}-1)^{2}+4\Lambda ^{2}\Omega
^{2}}{(\Delta ^{2}-\Lambda ^{2})^{2}+4\Lambda ^{2}\Omega ^{2}}}\Big],
\end{array}
\label{eq23}
\end{equation}
\begin{equation}
\begin{array}{ll}
Y^{2}& =\left( \frac{c}{V}\right) ^{2}=\frac{\epsilon _{0}}{2s^{2}}\Big[%
-(1-s^{2})+\frac{(1-s^{2})(\Delta ^{2}-\Lambda ^{2})+4\Lambda ^{2}}{(\Delta
^{2}-\Lambda ^{2})^{2}+4\Lambda ^{2}\Omega ^{2}} \\
& +(1+s^{2})\sqrt{\frac{(\Delta ^{2}-\Lambda ^{2}-1)^{2}+4\Lambda ^{2}\Omega
^{2}}{(\Delta ^{2}-\Lambda ^{2})^{2}+4\Lambda ^{2}\Omega ^{2}}}\Big].
\end{array}
\label{eq24}
\end{equation}
In the limit of a uniform wave ($\tau \rightarrow \infty $), when $\Lambda
^{2}\rightarrow 0,\,s^{2}\rightarrow 0$, Eq.~(\ref{eq23}) reproduces the
dispersion law of linear plane waves,
\begin{equation}
\left( \frac{ck}{\omega }\right) ^{2}=\epsilon _{0}\left( 1-\frac{1}{\Delta
^{2}}\right) \geq 0,  \label{eq25}
\end{equation}
which after substitution of Eq.~(\ref{eq20}) can be transformed into the
standard form. To find the velocity of the envelope of a linear wave, we
have to take the limit $s^{2}\rightarrow 0$ of Eq.~(\ref{eq24}) which gives
\begin{equation}
\left( \frac{c}{V}\right) ^{2}=\frac{\epsilon _{0}}{\Delta ^{6}(\Delta
^{2}-1)}\left[ \Delta ^{2}(\Delta ^{2}-1)+\Omega ^{2}\right] ^{2}.
\label{eq26}
\end{equation}
As one could expect, this velocity $V$ coincides with the group velocity of
propagating linear plane waves with the dispersion law (\ref{eq4}).

In Eqs.~(\ref{eq23}) and (\ref{eq24}) three parameters, $\Delta ^{2}$, $%
s^{2} $, and $\Omega ^{2}$, depend on the frequency $\omega $. For further
investigation it is convenient to express $s^{2}$ and $\Omega ^{2}$ in terms
of $\Delta ^{2}$,
\begin{equation}
\Omega ^{2}=\Delta ^{2}+\kappa ^{2},\quad s^{2}=\frac{\Lambda ^{2}}{\Delta
^{2}+\kappa ^{2}},\quad \kappa ^{2}=\frac{\omega _{T}^{2}}{\omega
_{L}^{2}-\omega _{T}^{2}},  \label{eq27}
\end{equation}
and then equations (\ref{eq23}) and (\ref{eq24}) take the form
\begin{equation}
\begin{array}{ll}
\left( \frac{ck}{\omega }\right) ^{2}& =\frac{1}{2}\Bigg[1-\frac{\Lambda ^{2}%
}{\Delta ^{2}+\kappa ^{2}}-\frac{\left( 1-\Lambda ^{2}/(\Delta ^{2}+\kappa
^{2})\right) (\Delta ^{2}-\Lambda ^{2})+4\Lambda ^{2}}{(\Delta ^{2}-\Lambda
^{2})^{2}+4\Lambda ^{2}(\Delta ^{2}+\kappa ^{2})} \\
& +\left( 1+\frac{\Lambda ^{2}}{\Delta ^{2}+\kappa ^{2}}\right) \sqrt{\frac{%
(\Delta ^{2}-\Lambda ^{2}-1)^{2}+4\Lambda ^{2}(\Delta ^{2}+\kappa ^{2})}{%
(\Delta ^{2}-\Lambda ^{2})^{2}+4\Lambda ^{2}(\Delta ^{2}+\kappa ^{2})}}\Bigg],
\end{array}
\label{eq28}
\end{equation}
\begin{equation}
\begin{array}{ll}
\left( \frac{c}{V}\right) ^{2}=& \frac{\Delta ^{2}+\kappa ^{2}}{2\Lambda ^{2}%
}\Bigg[-1+\frac{\Lambda ^{2}}{\Delta ^{2}+\kappa ^{2}}+\frac{\left(
1-\Lambda ^{2}/(\Delta ^{2}+\kappa ^{2})\right) (\Delta ^{2}-\Lambda
^{2})+4\Lambda ^{2}}{(\Delta ^{2}-\Lambda ^{2})^{2}+4\Lambda ^{2}(\Delta
^{2}+\kappa ^{2})} \\
& +\left( 1+\frac{\Lambda ^{2}}{\Delta ^{2}+\kappa ^{2}}\right) \sqrt{\frac{%
(\Delta ^{2}-\Lambda ^{2}-1)^{2}+4\Lambda ^{2}(\Delta ^{2}+\kappa ^{2})}{%
(\Delta ^{2}-\Lambda ^{2})^{2}+4\Lambda ^{2}(\Delta ^{2}+\kappa ^{2})}}\Bigg],
\end{array}
\label{eq29}
\end{equation}
where we have also put $\epsilon _{0}=1$ that is equivalent to the
replacement $c\rightarrow c/\sqrt{\epsilon _{0}}$. We remind that in Eqs. (%
\ref{eq28}) and (\ref{eq29}) $\kappa ^{2}$ is a constant determined by the
system under consideration, the parameter $\Delta ^{2}$ measures the
frequency of the wave and the parameter $\Lambda ^{2}$ measures the duration
of the pulse.

Linear uniform waves cannot propagate with frequency within the polariton
gap (\ref{eq5}), or
\begin{equation}
0<\Delta ^{2}<1.  \label{eq29a}
\end{equation}
However, at finite values of $\tau $, two branches of the dispersion curve
are joined into one curve. Plots of $\Delta ^{2}$ against $(ck/\omega )^{2}$
at several values of $\Lambda ^{2}$ are shown in Fig.~1. As we see, these
curves depend essentially on the values of $\Lambda ^{2}$, and only at $%
\Lambda \ll 1$ and far enough from the polariton gap we can apply the usual
approach with transition to NLS equation for the envelope function. The
velocity parameter $(c/V)^{2}$ of the pulse as a function of $\Delta ^{2}$
at several values of $\Lambda ^{2}$ is shown in Fig.~2. It has real values
even at frequencies inside the polariton gap (\ref{eq29a}). It is important
to note that if general nonlinear equations have a pulse solution of the
form (\ref{eq6})--(\ref{eq7}) , then its velocity must coincide with its
``linear approximation'' (\ref{eq29}) calculated for the tails of the pulse.

\section{Soliton solutions}

To find soliton solutions, we return to the exact equations (\ref{eq8})--(%
\ref{eq11}) replacing differentiation with respect to $\xi =t-x/V$ by
differentiation with respect to $\zeta =\xi /\tau $. Taking into account
Eqs.~(\ref{eq17})--(\ref{eq20}), we arrive at the system
\begin{equation}
\begin{array}{ll}
\Lambda ^{2}\ddot{u}-& (\Delta ^{2}+2\Lambda \sqrt{\Delta ^{2}+\kappa ^{2}}%
\dot{\phi}+\Lambda ^{2}\dot{\phi}^{2})u-2(\Lambda \sqrt{\Delta ^{2}+\kappa
^{2}}+\Lambda ^{2}\dot{\phi})\dot{v} \\
& -\Lambda ^{2}\ddot{\phi}v+\tilde{\chi}(u^{2}+v^{2})u=(1/4\pi )\mathcal{E},
\end{array}
\label{eq30}
\end{equation}
\begin{equation}
\begin{array}{ll}
\Lambda ^{2}\ddot{v}-(\Delta ^{2}+2\Lambda \sqrt{\Delta ^{2}+\kappa ^{2}}%
\dot{\phi}+\Lambda ^{2}\dot{\phi}^{2})v& +2(\Lambda \sqrt{\Delta
^{2}+\kappa ^{2}}+\Lambda ^{2}\dot{\phi})\dot{u} \\
& +\Lambda ^{2}\ddot{\phi}u+\tilde{\chi}(u^{2}+v^{2})v=0,
\end{array}
\label{eq31}
\end{equation}
\begin{equation}
\begin{array}{l}
 \Lambda ^{2}\left[ \left( \frac{c}{V}\right) ^{2}-1\right] \ddot{\mathcal{E%
}}-\Bigg\{(\Delta ^{2}+\kappa ^{2})\left[ \left( \frac{ck}{\omega }\right)
^{2}-1\right] \\
 +2\Lambda \sqrt{\Delta ^{2}+\kappa ^{2}}\left[ \frac{c^{2}k}{V\omega }-1%
\right] \dot{\phi}+\Lambda ^{2}\left[ \left( \frac{c}{V}\right) ^{2}-1\right]
\dot{\phi}^{2}\Bigg\}\mathcal{E} \\
 =4\pi \Big[ \Lambda ^{2}\ddot{u}-u(\sqrt{\Delta ^{2}+\kappa ^{2}}+\Lambda
\dot{\phi})^{2}\\
-\Lambda ^{2}v\ddot{\phi}-2(\Lambda \sqrt{\Delta ^{2}+\kappa
^{2}}+\Lambda ^{2}\dot{\phi})\dot{v}\Big] ,
\end{array}
\label{eq32}
\end{equation}
\begin{equation}
\begin{array}{l}
 \Lambda ^{2}\left[ \left( \frac{c}{V}\right) ^{2}-1\right] \ddot{\phi}{%
\mathcal{E}}+2\Big\{ \Lambda \sqrt{\Delta ^{2}+\kappa ^{2}}\left( \frac{%
c^{2}k}{V\omega }-1\right)\\
 +\Lambda ^{2}\left[ \left( \frac{c}{V}\right)
^{2}-1\right] \dot{\phi}\Big\} \dot{\mathcal{E}} \\
 =4\pi \Big[ \Lambda ^{2}\ddot{v}-v(\sqrt{\Delta ^{2}+\kappa ^{2}}+\Lambda
\dot{\phi})^{2}\\
+\Lambda ^{2}u\ddot{\phi}+2(\Lambda \sqrt{\Delta ^{2}+\kappa
^{2}}+\Lambda ^{2}\dot{\phi})\dot{u}\Big] ,
\end{array}
\label{eq33}
\end{equation}
where
\begin{equation}
\tilde{\chi}=\chi /(\omega _{L}^{2}-\omega _{T}^{2}),
\end{equation}
and overdot stands now for derivative with respect to $\zeta =\xi /\tau $.

Let us consider long pulses with $\Lambda ^{2}\ll \Delta ^{2}$. To this end,
we introduce a small parameter $\varepsilon $ by
\begin{equation}
\Lambda ^{2}=\varepsilon ^{2}\Delta ^{2},\qquad \varepsilon \ll 1,
\label{eq34}
\end{equation}
so that Eqs.~(\ref{eq30})--(\ref{eq33}) can be expanded in powers of $%
\varepsilon $. Since Eqs.~(\ref{eq28}) and (\ref{eq29}) lead to different
series expansions in different intervals of $\Delta ^{2}$, we have to
consider all these cases separately.

\subsection{Long pulse above the polariton gap}

We begin with the case of waves with
\begin{equation}
\Delta ^{2}>1,  \label{eq35}
\end{equation}
for which a nonlinear pulse can be presented as the envelope of propagating
linear waves with dispersion law (\ref{eq25}). For long pulses with
\begin{equation}
\varepsilon ^{2}\ll 1,\qquad \varepsilon ^{2}\ll \Delta ^{2}-1  \label{eq36}
\end{equation}
the coefficients in Eqs.~(\ref{eq32}), (\ref{eq33}) can be represented as
series expansions in powers of $\varepsilon $:
\begin{equation}
\left( \frac{ck}{\omega }\right) ^{2}-1=-\frac{1}{\Delta ^{2}}+\frac{\Delta
^{4}-3\kappa ^{2}+4\Delta ^{2}\kappa ^{2}}{\Delta ^{4}(\Delta ^{2}-1)}%
\varepsilon ^{2}+\ldots =-\frac{1}{\Delta ^{2}}+\alpha _{2}\varepsilon
^{2}+\ldots ,  \label{eq37}
\end{equation}
\begin{equation}
\Lambda ^{2}\left[ \left( \frac{c}{V}\right) ^{2}-1\right] =\frac{\Delta
^{6}+2\Delta ^{4}\kappa ^{2}+\kappa ^{4}}{\Delta ^{4}(\Delta ^{2}-1)}%
\varepsilon ^{2}+\ldots =\gamma _{2}\varepsilon ^{2}+\ldots ,  \label{eq38}
\end{equation}
\begin{equation}
2\Lambda \left( \frac{c^{2}k}{V\omega }-1\right) =\frac{2\kappa ^{2}}{\Delta
^{3}}\varepsilon -\frac{4\kappa ^{2}(\Delta ^{2}+2\kappa ^{2})}{\Delta ^{7}}%
\varepsilon ^{3}+\ldots =\beta _{1}\varepsilon -\beta _{3}\varepsilon
^{3}+\ldots .  \label{eq39}
\end{equation}
We suppose that functions $\mathcal{E}$, $\dot{\phi}$, $u$, $v$ can also be
represented in the form of series expansions, $\mathcal{E=}\varepsilon ^{\nu
}\sum_{n=0}^{\infty }\varepsilon ^{n}\mathcal{E}_{n}$ \ and in the same way for $%
\dot{\phi}$, $u$, $v$. Analysis of equations (\ref{eq30})--(\ref{eq33}) and (%
\ref{eq37})--(\ref{eq39}) shows that for self-consistency of the procedure
the series expansions of the fields must be as follows:
\begin{equation}
\begin{array}{ll}
\mathcal{E}& =\varepsilon (\mathcal{E}_{0}+\varepsilon ^{2}\mathcal{E}%
_{2}+\ldots ),\quad \dot{\phi}=\varepsilon (\theta _{0}+\varepsilon
^{2}\theta _{2}+\ldots ), \\
u& =\varepsilon (u_{0}+\varepsilon ^{2}u_{2}+\ldots ),\quad v=\varepsilon
(\varepsilon v_{1}+\varepsilon ^{3}v_{3}+\ldots ).
\end{array}
\label{eq40}
\end{equation}
So substitution of these expansions into Eqs.~(\ref{eq30})--(\ref{eq33})
yields a sequence of equations for the coefficients $\mathcal{E}_{0},\,%
\mathcal{E}_{2},\,\theta _{0},\ldots $. In the first approximation we obtain
the relations
\begin{equation}
u_{0}=-\frac{1}{4\pi \Delta ^{2}}\mathcal{E}_{0},\quad v_{1}=\frac{2\sqrt{%
\Delta ^{2}+\kappa ^{2}}}{\Delta }\dot{u}_{0},  \label{eq41}
\end{equation}
which correspond to the plane wave solution with constant amplitude. In the
next approximation we obtain from Eqs.~(\ref{eq30}) and (\ref{eq32})
\begin{equation}
\begin{array}{l}
 -\ddot{\mathcal{E}}_{0}+4\frac{\sqrt{\Delta ^{2}+\kappa ^{2}}}{\Delta ^{2}}%
\ddot{\mathcal{E}}_{0}-\frac{\tilde{\chi}}{16\pi ^{2}\Delta ^{6}}\mathcal{E}%
_{0}^{3}=\mathcal{E}_{2}+4\pi \Delta ^{2}u_{2}-\frac{2\sqrt{\Delta
^{2}+\kappa }}{\Delta }\theta _{0}\mathcal{E}_{0}, \\
 \left( \gamma _{2}+1-4\frac{\sqrt{\Delta ^{2}+\kappa ^{2}}}{\Delta ^{2}}%
\right) \ddot{\mathcal{E}}_{0}-(\Delta ^{2}+\kappa ^{2})\alpha _{2}\mathcal{E%
}_{0} \\
 =-\frac{\Delta ^{2}+\kappa ^{2}}{\Delta ^{2}}\left( \mathcal{E}_{2}+4\pi
\Delta ^{2}u_{2}-\frac{2\sqrt{\Delta ^{2}+\kappa }}{\Delta }\theta _{0}%
\mathcal{E}_{0}\right) .
\end{array}
\end{equation}
Combination of these two equations yields after simple transformations the
equation for $\mathcal{E}_{0}$:
\begin{equation}
\ddot{\mathcal{E}}_{0}=\mathcal{E}_{0}+\frac{\tilde{\chi}}{16\pi ^{2}\Delta
^{8}\alpha _{2}}\mathcal{E}_{0}^{3}.  \label{eq42}
\end{equation}
If $\tilde{\chi}<0$, then it has the soliton solution
\begin{equation}
\mathcal{E}_{0}=\frac{a}{\cosh \zeta },  \label{eq43}
\end{equation}
where
\begin{equation}
a=\sqrt{\frac{32\pi ^{2}\Delta ^{8}\alpha _{2}}{|\tilde{\chi}|}},\quad
\alpha _{2}=\frac{\Delta ^{4}-3\kappa ^{2}+4\Delta ^{2}\kappa ^{2}}{\Delta
^{4}(\Delta ^{2}-1)}  \label{eq44}
\end{equation}
and
\begin{equation}
\zeta =\frac{1}{\tau }\left( t-\frac{x}{V}\right) ,\quad \left( \frac{c}{V}%
\right) ^{2}=\frac{(\Delta ^{4}+\kappa ^{2})^{2}}{\Delta ^{6}(\Delta ^{2}-1)}%
;  \label{eq45}
\end{equation}
this shows that $V$ is equal to the group velocity given by Eq.~(\ref{eq26})
and does not depend on the duration of the pulse. Since $\varepsilon
=\Lambda /\Delta =1/(\sqrt{\omega ^{2}-\omega _{T}^{2}}\,\tau ),$ $\tilde{%
\chi}=\chi /(\omega _{L}^{2}-\omega _{T}^{2}),$ we obtain
\begin{equation}
\mathcal{E}=\sqrt{\frac{32\pi ^{2}\Delta ^{6}\alpha _{2}}{|\chi |}}\,\frac{1%
}{\tau }\frac{1}{\cosh \left[ \frac{1}{\tau }\left( t-\frac{x}{V}\right) %
\right] }.  \label{eq46}
\end{equation}
This is the NLS type soliton solution.

Note that the above calculations were made for arbitrary values of $\kappa$.
However, in majority of applications we have $(\om_L-\om_T)\ll\om_T$,
that is
\begin{equation}\label{aa}
  \kappa^2=\frac{\om_T^2}{\om_L^2-\om_T^2}\simeq\frac{\om_T}{2(\om_L-\om_T)}
  \gg1.
\end{equation}
Then expansions (\ref{eq37})-(\ref{eq39})  are valid only under the condition
that
\begin{equation}\label{bb}
  \kappa^2\eps^2\simeq 1/4(\om_L-\om_T)^2\tau^2\ll 1,
\end{equation}
which means that the spectral width of the pulse $\sim1/\tau$ is much less
than the width of the gap $(\om_L-\om_T)$. In fact, the condition (\ref{bb}) is
satisfied well enough already for $1/\tau\simeq (\om_L-\om_T)/2$.

\subsection{Long pulse at $\Delta^2=1$}

In this case a pulse cannot be described by the NLS equation
for the wave packet of waves with wave vectors around some nonzero
value. From Eqs.~(\ref{eq28})-(\ref{eq29}) we have the expansions
\begin{equation}
\left( \frac{ck}{\omega }\right) ^{2}-1=-1+\sqrt{1+\kappa ^{2}}\varepsilon
+\ldots =-1+\bar{\alpha}_{1}\varepsilon +\ldots ,  \label{eq47}
\end{equation}
\begin{equation}
\begin{array}{ll}
\Lambda ^{2}\left[ \left( \frac{c}{V}\right) ^{2}-1\right] &=(1+\kappa
^{2})^{3/2}\varepsilon -(2\kappa ^{4}+\frac{3}{2}\kappa ^{2}+\frac{1}{2}%
)\varepsilon ^{2}+\ldots\\
 &=\bar{\gamma}_{1}\varepsilon -\bar{\gamma}%
_{2}\varepsilon ^{2}+\ldots ,
\end{array} \label{eq48}
\end{equation}
\begin{equation}
2\Lambda \left( \frac{c^{2}k}{V\omega }-1\right) =2\kappa ^{2}\varepsilon
-4(1+2\kappa ^{2})\kappa ^{2}\varepsilon ^{3}+\ldots =\bar{\beta}%
_{1}\varepsilon -\bar{\beta}_{3}\varepsilon ^{3}+\ldots ,  \label{eq49}
\end{equation}
Again we look for the solution of Eqs.~(\ref{eq30})--(\ref{eq33}) in the
form of series expansions which in this case are
\begin{equation}
\begin{array}{ll}
\mathcal{E}=& \varepsilon ^{1/2}(\mathcal{E}_{0}+\varepsilon \mathcal{E}%
_{1}+\ldots ),\quad \dot{\phi}=\varepsilon ^{1/2}\theta _{0}+\ldots , \\
u=& \varepsilon ^{1/2}(u_{0}+\varepsilon u_{1}+\ldots ),\quad v=\varepsilon
^{3/2}v_{1}+\ldots .
\end{array}
\label{eq50}
\end{equation}
In the first approximation we have
\begin{equation}
u_{0}=-\mathcal{E}_{0}/4\pi ,\quad v_{1}=2\sqrt{1+\kappa ^{2}}\,\dot{u}_{0}.
\label{eq51}
\end{equation}
In the next approximation Eqs.~(\ref{eq30}) and (\ref{eq33}) give
\begin{equation}
-u_{1}+\tilde{\chi}u_{0}^{3}=\mathcal{E}_{1}/4\pi ,\quad (1+\kappa
^{2})^{1/2}\ddot{\mathcal{E}}_{0}+\mathcal{E}_{1}-\bar{\alpha}_{1}\mathcal{E}%
_{0}=-4\pi u_{1}.
\end{equation}
Hence,
\begin{equation}
\mathcal{E}_{1}+4\pi u_{1}=-\frac{\tilde{\chi}}{16\pi ^{2}}\mathcal{E}%
_{0}^{3},\quad (1+\kappa ^{2})^{1/2}\ddot{(\mathcal{E}}_{0}-\mathcal{E}%
_{0})=-(\mathcal{E}_{1}+4\pi u_{1}),
\end{equation}
and we arrive at the equation for $\mathcal{E}_{0}$:
\begin{equation}
\ddot{\mathcal{E}}_{0}=\mathcal{E}_{0}+\frac{\tilde{\chi}}{16\pi
^{2}(1+\kappa ^{2})^{1/2}}\mathcal{E}_{0}^{3}.  \label{eq52}
\end{equation}
Thus, we obtain the soliton solution
\begin{equation}
\mathcal{E}=\frac{\varepsilon ^{1/2}a}{\cosh \left[ \frac{1}{\tau }\left( t-%
\frac{x}{V}\right) \right] },\quad a=\sqrt{\frac{32\pi ^{2}(1+\kappa
^{2})^{1/2}}{|\tilde{\chi}|}},  \label{eq53}
\end{equation}
where $V$ is given by
\begin{equation}
\left( \frac{c}{V}\right) ^{2}=1+\frac{(1+\kappa ^{2})^{3/2}}{\varepsilon }.
\label{eq54}
\end{equation}
Taking into account that $\varepsilon =\Lambda =1/\left( \sqrt{\omega
_{L}^{2}-\omega _{T}^{2}}\,\cdot \tau \right) $, we can rewrite Eqs.~(\ref
{eq53}), (\ref{eq54}) in terms of the physical parameters
\begin{equation}
\mathcal{E}=\sqrt{\frac{32\pi ^{2}\omega _{L}}{|\chi |\tau }}\cdot \frac{1}{%
\cosh \left[ \frac{1}{\tau }\left( t-\frac{x}{V}\right) \right] },\qquad
\left( \frac{c}{V}\right) ^{2}=1+\frac{\omega _{L}^{3}\tau }{\omega
_{L}^{2}-\omega _{T}^{2}}.  \label{eq54a}
\end{equation}
Thus the velocity of the pulse depends on $\tau $ (curves in Fig.~2
intersect the straight line $\Delta ^{2}=1$ at different points depending on
$\tau $). Although the parameter $\kappa$ disappeared from Eq.~(\ref{eq54a}),
the expansions (\ref{eq47})-(\ref{eq49}) are valid for $\kappa\gg 1$ provided
the inequality $(\kappa\eps)^2\ll1$ is satisfied. The ratio of the amplitude
of the solution (\ref{eq54a}) to
that of the solution (\ref{eq46}) is of order of magnitude
$$\sim(\kappa\eps)^{-1/2}\sim\sqrt{(\om_L-\om_T)\tau}\gg 1,$$ that is the
amplitude at the boundary of the gap is
much greater  than the amplitude of the soliton solution far enough from
the gap. This means that the pulse should be intensive enough to deform
the gap to such extent that propagation of the wave with frequency
$\Delta^2=1$ becomes possible. Beyond the gap there are linear waves
which can propagate with arbitrarily small amplitudes  and nonlinear effects
must only compensate dispersive
spreading of the wave packet built from linear waves.

\subsection{Long pulse inside the polariton gap}

For frequencies inside the polariton gap,
\begin{equation}
0<\Delta ^{2}<1,  \label{eq55}
\end{equation}
we have the series expansions
\begin{equation}
\left( \frac{ck}{\omega }\right) ^{2}-1=-1+\frac{(\Delta ^{4}+\kappa
^{2})^{2}}{\Delta ^{4}(\Delta ^{2}-1)(\Delta ^{2}+\kappa ^{2})}\varepsilon
^{2}+\ldots =-1+\tilde{\alpha}_{2}\varepsilon ^{2}+\ldots ,  \label{eq56}
\end{equation}
\begin{equation}
\begin{array}{ll}
\Lambda ^{2}& \left[ \left( \frac{c}{V}\right) ^{2}-1\right] =\frac{%
(1-\Delta ^{2})(\Delta ^{2}+\kappa ^{2})}{\Delta ^{2}} \\
& +\frac{\Delta ^{8}+5\Delta ^{4}\kappa ^{2}-3\kappa ^{4}+\Delta ^{2}\kappa
^{2}(4\kappa ^{2}-3)}{\Delta ^{4}(1-\Delta ^{2})}\varepsilon ^{2}+\ldots =%
\tilde{\gamma}_{0}+\tilde{\gamma}_{2}\varepsilon ^{2}+\ldots ,
\end{array}
\label{eq57}
\end{equation}
\begin{equation}
2\Lambda \left( \frac{c^{2}k}{V\omega }-1\right) =\frac{2\kappa ^{2}}{\Delta
^{3}}\varepsilon -\frac{4\kappa ^{2}(\Delta ^{2}+2\kappa ^{2})}{\Delta ^{7}}%
\varepsilon ^{3}+\ldots =\beta _{1}\varepsilon -\beta _{3}\varepsilon
^{3}+\ldots ,  \label{eq58}
\end{equation}
where
\begin{equation}
\varepsilon \ll (1-\Delta ^{2}),\quad \varepsilon \ll \Delta ^{2}.  \label{eq59}
\end{equation}
Because $\tilde{\gamma}_{0}\neq 0$, the soliton solution is obtained in the
first approximation, so that $\mathcal{E}$ and $u$ do not have a small
factor proportional to some power of $\varepsilon $. The equations of the
first approximation read
\begin{equation}
-\Delta ^{2}u+\tilde{\chi}u^{3}=(1/4\pi )\mathcal{E},  \label{eq60}
\end{equation}
\begin{equation}
\frac{1-\Delta ^{2}}{\Delta ^{2}}\ddot{\mathcal{E}}+\mathcal{E}=-4\pi u.
\label{eq61}
\end{equation}
If we were able to express $u$ in terms of $\mathcal{E}$ from Eq.~(\ref{eq60}%
) and substitute the result into Eq.~(\ref{eq61}), then we would obtain the
equation for $\mathcal{E}$ having solitary wave solutions. Unfortunately,
that can be done only numerically except for the case when
\begin{equation}
|\tilde{\chi}|u^{2}\ll \Delta ^{2}.  \label{eq62}
\end{equation}
In this limit we have
\begin{equation}
u\cong -\frac{1}{4\pi \Delta ^{2}}\mathcal{E}-\frac{\tilde{\chi}}{\Delta
^{2}(4\pi \Delta ^{2})^{3}}\mathcal{E}^{3}  \label{eq63}
\end{equation}
and Eq.~(\ref{eq61}) takes the form
\begin{equation}
\ddot{\mathcal{E}}=\mathcal{E}+\frac{\tilde{\chi}}{16\pi ^{2}\Delta
^{6}(1-\Delta ^{2})}\mathcal{E}^{3}.  \label{eq64}
\end{equation}
It has the soliton solution
\begin{equation}
\mathcal{E}=\sqrt{\frac{32\pi ^{2}\Delta ^{6}(1-\Delta ^{2})(\om_L^2-\om_T^2)}{|{%
\chi}|}}\frac{1}{\cosh \left[ \frac{1}{\tau }\left( t-\frac{x}{V}\right) %
\right] },  \label{eq65}
\end{equation}
where the pulse velocity is given by
\begin{equation}
\left( \frac{c}{V}\right) ^{2}=1+\frac{(1-\Delta ^{2})(\Delta ^{2}+\kappa
^{2})(\omega ^{2}-\omega _{T}^{2})\tau ^{2}}{\Delta ^{2}}.
\label{eq66}
\end{equation}
The condition (\ref{eq62}) can be transformed with the use of the estimate $%
u\sim a/(4\pi \Delta ^{2})$ into
\begin{equation}
1-\Delta ^{2}\ll 1,  \label{eq67}
\end{equation}
that is the frequency must be close enough to the upper limit of the
polariton gap.

The solution (\ref{eq65}) only applies to long enough pulses
\begin{equation}
\varepsilon ^{2}\ll 1-\Delta ^{2}.
\end{equation}
Note that in this case the amplitude does not depend on the pulse width $%
\tau $, but there is a strong dependence of its velocity on $\tau $.
If $\kappa\gg 1$, then the ratio of the amplitude (\ref{eq65}) to the
amplitude (\ref{eq46}) is of order of magnitude
$$\sim(\kappa\eps)^{-1}\sim(\om_L-\om_T)\tau\gg 1,$$
i.e.,  it is much greater, as one could expect, than the amplitude
of the soliton solution (\ref{eq54a}) at the boundary of the gap.

When the condition (\ref{eq67}) is not fulfilled, we have to solve Eqs.~(\ref
{eq60}), (\ref{eq61}) numerically. To this end, let us introduce new
variables
\begin{equation}
\mathcal{E}=E/\sqrt{|\widetilde{\chi }|},\qquad u=U/\sqrt{|\widetilde{\chi }|%
},  \label{a1}
\end{equation}
where $\widetilde{\chi }=-|\widetilde{\chi }|$, so that Eqs.~(\ref{eq60},\ref
{eq61}) reduce to
\begin{equation}
\Delta ^{2}U+U^{3}=-(1/4\pi )E,\qquad \frac{1-\Delta ^{2}}{\Delta ^{2}}\ddot{%
E}+E=-4\pi U.  \label{a2}
\end{equation}
Numerical solution of the first equation (\ref{a2}) yields the function
\begin{equation}
U=U(E)  \label{a3}
\end{equation}
the substitution of which into the second equation of (\ref{a2}) gives
the differential equation for $E$:
\begin{equation}
\frac{1-\Delta ^{2}}{\Delta ^{2}}\frac{d^{2}E}{d\zeta ^{2}}=-E-4\pi U(E)
\label{a4}
\end{equation}
or
\begin{equation}
\frac{d^{2}E}{d\zeta ^{2}}=-\frac{\partial V(E)}{\partial E},  \label{a5}
\end{equation}
where
\begin{equation}
V(E)=\frac{\Delta ^{2}}{1-\Delta ^{2}}\left( \frac{1}{2}E^{2}+4\pi
\int_{0}^{E}U(E^{\prime })dE^{\prime }\right)  \label{a6}
\end{equation}
may be considered as a potential in which a particle moves according to the
Newton equation (\ref{a5}). A typical plot of the potential (\ref{a6}) is
shown in Fig.~3. The maximum value of $E$ (soliton's amplitude $a$) is
determined by the point $E=a$ where $V(E)$ vanishes. Then the solution of
Eq.~(\ref{a4}) with initial conditions $E(0)=a,\,\,E^{\prime }(0)=0$
provides the soliton solution. We have done these calculations for several
values of $\Delta $. The results are shown in Fig.~4 where the dependence of
the pulse profile $E$ on the scaled variable $\sqrt{\Delta ^{2}/(1-\Delta
^{2})}\,\zeta $ is plotted. The dependence of the amplitude on $\Delta $ is
shown in Fig.~5.\bigskip

\section{Conclusion}

In conclusion, the nonlinear optical pulse propagation in the frequency
region in the vicinity of polariton gap is studied. The problem is described
by the coupled set of the Maxwell equations for electromagnetic field and
material equation for macroscopic polarization allowing for Kerr-like
nonlinearity. To solve this nonlinear system of equations analytically, the
approach ~\cite{AI} based on series expansion in powers of a small parameter
related to the width of the polariton gap and pulse duration is used.
Different bright solitary wave solutions, depending on the position of
carrier wave frequency with respect to the polariton gap, are found and
their parameters are expressed in terms of material system parameters.
Outside the polariton gap the soliton solution corresponds to well-known
soliton of the NLS equation for the envelope of the wave packet made of
plane waves. However, inside the polariton gap there are no plane wave
solutions and  the notion of their envelope looses its physical sense.
Nonetheless, there the solitary wave solutions are possible with high
enough values of electromagnetic field strength so that local value of
the polariton gap diminishes in the center of the polariton gap due to
Kerr nonlinearity. The difference in physical situations outside and
inside the polariton gap is reflected in different dependencies of
soliton's amplitude on the pulse duration $\tau$---the amplitude does not
depend on $\tau$ inside the gap, it is proportional to $\tau^{-1/2}$ at
the gap boundary and proportional to $\tau^{-1}$ far enough from the gap.

S.D. thanks Marseilles University for kind hospitality and
NATO Linkage Grant PST. CLG. 978177 for partial support.


\newpage

\centerline{\bf Figure captions }

\bigskip

Fig. 1. Dispersion relation of the carrier wave for different values of the
pulse duration $\tau $ measured by the parameter $\Lambda$
(see Eq.~(\ref{eq18})).

\bigskip

Fig.~2. Pulse velocity as a function of the carrier wave frequency for
different values of the pulse duration $\tau $ measured by the parameter $\Lambda$
(see Eq.~(\ref{eq18})).

\bigskip

Fig.~3. Potential $V$ as a function of the electric field amplitude for $%
\Delta=0.2$.

\bigskip

Fig. 4. Profiles of solitary wave solutions for three different values of
the carrier wave frequency (see Eq.~(\ref{eq20})) inside the polariton gap.

\bigskip

Fig. 5. Dependence of the amplitude of the solitary wave on the carrier wave
frequency (see Eq.~(\ref{eq20})) inside the polariton gap.

\bigskip

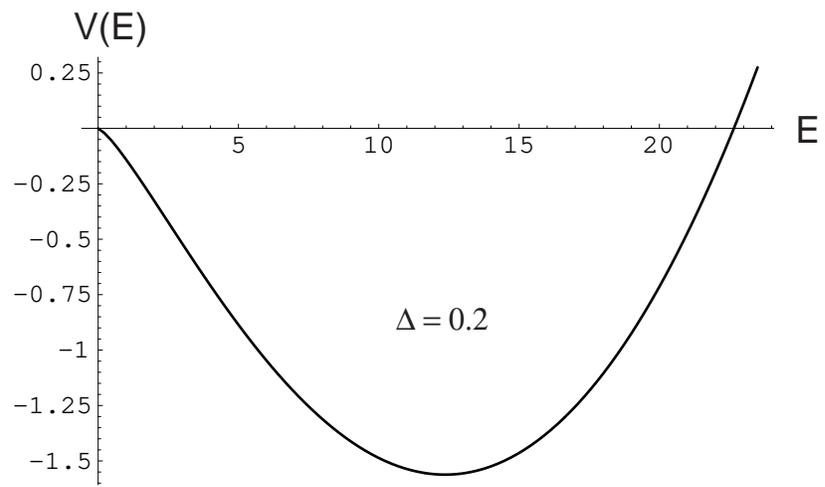
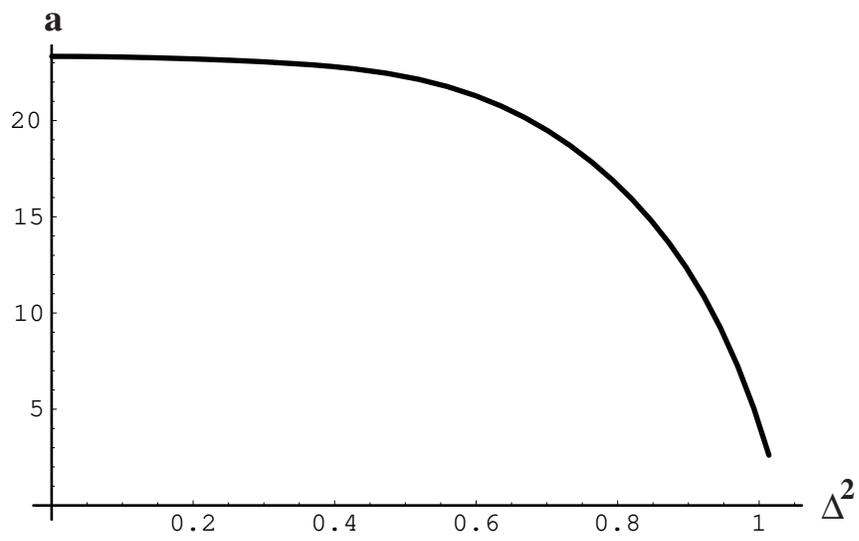
\end{document}